\documentclass[12pt,titlepage]{article}
\usepackage{amsfonts}
\usepackage{geometry}
\usepackage{times}
\usepackage[onehalfspacing]{setspace}

\input{tcilatex}
\begin{document}

\title{Analysis of AneuRisk65 data: warped logistic discrimination}
\author{Daniel Gervini \\
Department of Mathematical Sciences, University of \\
Wisconsin--Milwaukee}
\maketitle

\begin{abstract}
We analyze the AneuRisk65 curvature functions using a likelihood-based
warping method for sparsely sampled curves, and combine it with logistic
regression in order to discriminate subjects with aneurysms at or after the
terminal bifurcation of the internal carotid artery (the most
life-threatening) from subjects with no aneurysms or aneurysms along the
carotid artery (the less serious). Significantly lower misclassification
rates are obtained when the warping functions are included in the logistic
discrimination model, rather than being treated as mere nuisance parameters.

\end{abstract}

\section{Introduction}

The organizers of this section of the workshop are to be congratulated for
their choice of data. Without being overly complicated, the AneuRisk65 data
(Sangalli \emph{et al.}, 2013) presents many non-trivial challenges for
analysis. For example: the 65 angiographic images are misaligned due to the
different placement of the patients with respect to the image-capturing
device; the images have different lengths, with the origin corresponding to
a physiologically recognizable landmark but the endpoints being arbitrary;
and the main feature of interest, the syphon (Piccinelli \emph{et al.},
2011), varies in shape and location from person to person.

My analysis of the data was done on the curvature functions, not on the 3D
images themselves; this avoids the problem of rotating and translating the
3D curves to remove subject-placement artifacts, but does not remove the
inherent variability in shape and location of the artery syphon,
corresponding to the peaks around $t=-40$ and $t=-20$ in Figure \ref%
{fig:data_curves} (the variable $t$ is negative arc length in this
parametrization, so the curves run \textquotedblleft
backwards\textquotedblright ). The problem of unequal endpoints is also
present whether we analyze the original 3D images or the one-dimensional
curvature functions. My approach here is to treat the shorter curves as
incomplete curves (which they are). Since the curves for patients with
aneurysms at or after the terminal bifurcation of the internal carotid
artery (the \textquotedblleft upper\textquotedblright\ group) rarely extend
beyond $t=-80$ (i.e.~the data is \emph{not }missing at random), we truncated
the curves at $t=-80$ in order to avoid artifacts. But many curves were
shorter than this, so the problem of unequal endpoints persists; we deal
with this by introducing a model that can handle missing data, as explained
below.

\FRAME{ftbpFU}{6.3175in}{4.8239in}{0pt}{\Qcb{Curvature functions,
down-sampled to 30 measurements per curve, for (a) the \textquotedblleft
upper\textquotedblright\ group of patients and (c) the \textquotedblleft
lower\textquotedblright\ and no-aneurysm groups of patients. The
corresponding warped curves are shown in (b) for the \textquotedblleft
upper\textquotedblright\ group and in (d) for the \textquotedblleft
lower\textquotedblright\ and no-aneurysm groups.}}{\Qlb{fig:data_curves}}{%
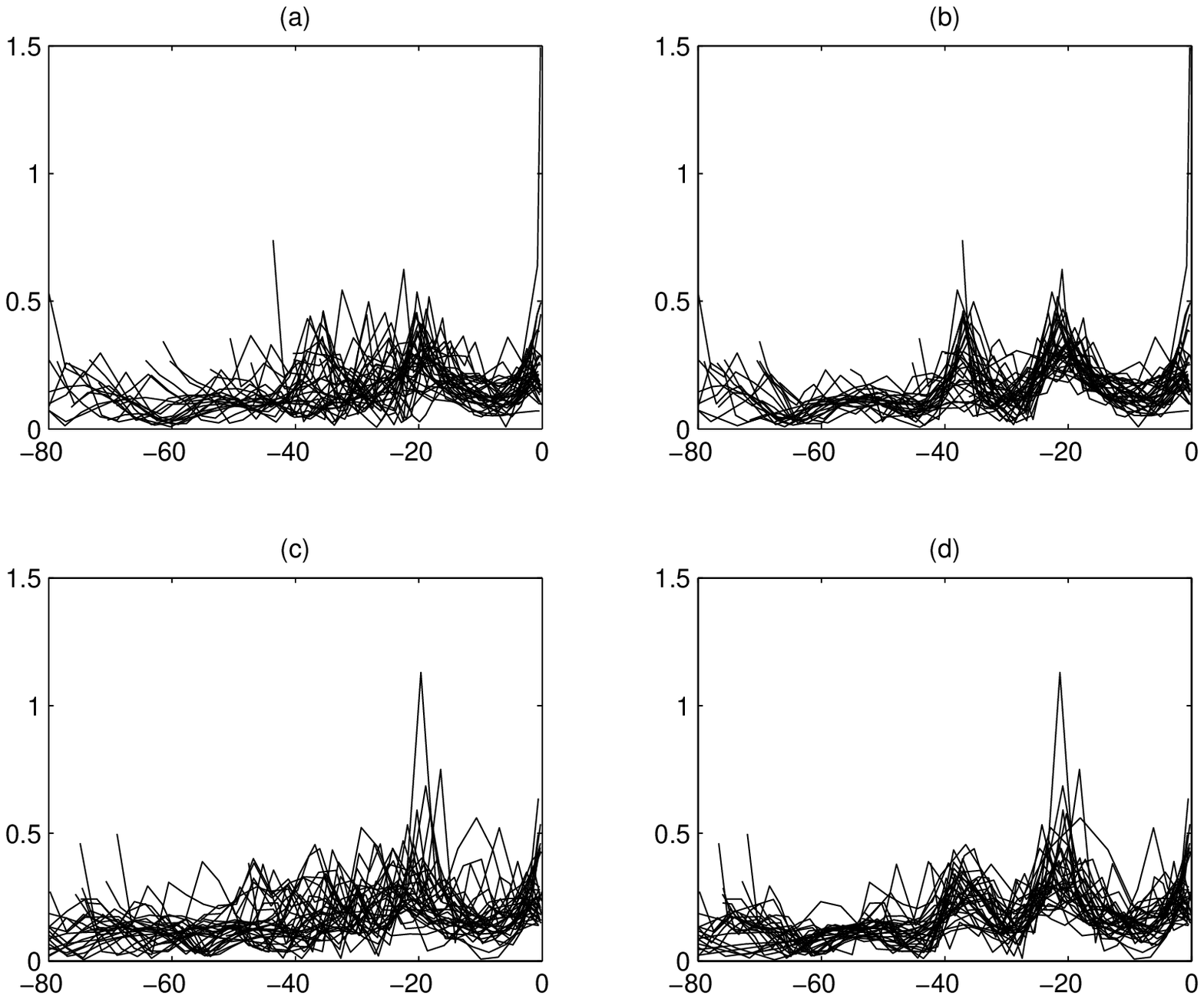}{\special{language "Scientific Word";type
"GRAPHIC";maintain-aspect-ratio TRUE;display "USEDEF";valid_file "F";width
6.3175in;height 4.8239in;depth 0pt;original-width 7.862in;original-height
5.9949in;cropleft "0";croptop "1";cropright "1";cropbottom "0";filename
'curves.eps';file-properties "XNPEU";}}

\section{The model}

Let $f_{1},\ldots ,f_{n}$ be the (complete, unobserved) curvature functions, 
$f_{i}:I\rightarrow \mathbb{R}$ with $I=[-80,0]$. The data actually observed
is of the form 
\begin{equation}
y_{ij}=f_{i}(t_{ij})+\varepsilon _{ij},\ \ j=1,\ldots ,m_{i},\ \ i=1,\ldots
,n,  \label{eq:raw-data}
\end{equation}%
for different grids $\{t_{i1},\ldots ,t_{im_{i}}\}$ and random errors $%
\{\varepsilon _{ij}\}$ (the errors could be assumed to be zero because the
curves were pre-smoothed, but the $\varepsilon $s are still a useful slack
variable to capture the random variation not explained by model (\ref{eq:KL}%
) below). The variability in location of the syphon will be accounted for by
the warping functions $h_{i}:I\rightarrow I$. We assume, then, that 
\begin{equation}
f_{i}(t)=\tilde{f}_{i}\{h_{i}^{-1}(t)\},  \label{eq:decomp}
\end{equation}%
where $\tilde{f}_{1},\ldots ,\tilde{f}_{n}$ are functions\ that, loosely
speaking, possess only amplitude variability and can therefore be modeled
with a parsimonious principal-component decomposition, 
\begin{equation}
\tilde{f}_{i}(t)=\mu (t)+\sum_{k=1}^{p}z_{ik}\xi _{k}(t),  \label{eq:KL}
\end{equation}%
where the $\xi _{k}$s are orthonormal functions in $\mathbb{L}^{2}(I)$ and
the $z_{ik}$s are uncorrelated with decreasing variances. In fact, we will
assume $\mathbf{z}_{i}=(z_{i1},\ldots ,z_{ip})\sim N_{p}(\mathbf{0},\mathbf{%
\Lambda })$ with $\mathbf{\Lambda }=\mathrm{diag}(\lambda _{1},\ldots
,\lambda _{p})$ and $\lambda _{1}\geq \cdots \geq \lambda _{p}>0$. We will
denote by $\mathcal{F}$ the family of functions spanned by (\ref{eq:KL}),
generally referred to as \textquotedblleft the template\textquotedblright\
in the warping literature. The $\xi _{k}$s, $\lambda _{k}$s and $\mu $ will
be estimated from the data; we will assume $\mu $ and the $\xi _{k}$s are
spline functions, thus reducing the estimation problem to a common
multivariate problem: given e.g.~a B-spline basis $\{\phi _{1},\ldots ,\phi
_{q}\}$, we assume $\mu (t)=\sum_{k=1}^{q}a_{l}\phi _{l}(t)$ and $\xi
_{k}(t)=\sum_{l=1}^{q}c_{kl}\phi _{l}(t)$ for parameters $\mathbf{a}%
=(a_{1},\ldots ,a_{q})$ and $\mathbf{c}_{k}=(c_{k1},\ldots ,c_{kq})$ to be
estimated from the data.

For the warping functions $h_{i}$ we also specify a family of functions $%
\mathcal{H}$ that is parsimonious but flexible enough to accommodate phase
variability at the salient features of the curves. The family of monotone
interpolating Hermite splines (Fritsch and Carlson, 1980) is very convenient
to work with. Given a knot vector $\mathbf{\tau }_{0}$ of \textquotedblleft
locations on interest\textquotedblright\ (for example, $\mathbf{\tau }%
_{0}=(-60,-40,-20)$ in our case) and any $\mathbf{\tau }_{i}$ with monotone
increasing coordinates, there exists an $h_{i}\in \mathcal{H}$ such that $%
h_{i}(\mathbf{\tau }_{0})=\mathbf{\tau }_{i}$; this interpolating property
provides all the warping flexibility we want at the features of interest,
without increasing the dimension of $\mathcal{H}$ unnecessarily. The
monotonicity of Hermite splines is very easy to enforce for \emph{any} $%
\mathbf{\tau }_{i}$s; see Fritsch and Carlson (1980). The individual $%
\mathbf{\tau }_{i}$s could be either specified by the researcher (as in
landmark registration) or treated as unobservable random effects, as we will
do here. Since the coordinates of the $\mathbf{\tau }_{i}$s must be strictly
increasing and fall within the range $I$, it is more convenient to transform
them into unconstrained vectors $\mathbf{\theta }_{i}$ using e.g.~the Jupp
transform, and assume $\mathbf{\theta }_{i}\sim N_{r}(\mathbf{\theta }_{0},%
\mathbf{\Sigma })$ with $\mathbf{\theta }_{0}$ the Jupp transform of $%
\mathbf{\tau }_{0}$ and $\mathbf{\Sigma }$ a covariance matrix to be
estimated from the data. Therefore, our warping functions will be
parameterized as $h_{i}(t)=g(t,\mathbf{\theta }_{i})$ for a fixed function $%
g $ that depends only on $\mathbf{\tau }_{0}$ (its exact form does not
matter here).

A brief digression: the decomposition (\ref{eq:decomp}) is clearly not
unique; given any $f_{i}$ and any arbitrary monotone function $h_{i}$, one
can always define $\tilde{f}_{i}=f_{i}\circ h_{i}$ and then the
decomposition $f_{i}(t)=\tilde{f}_{i}\{h_{i}^{-1}(t)\}$ trivially follows.
So it does not make sense to talk about \textquotedblleft
the\textquotedblright\ warping component $h_{i}$ and \textquotedblleft
the\textquotedblright\ amplitude component $\tilde{f}_{i}$ for a given $%
f_{i} $. Nevertheless, for a \emph{given} template $\mathcal{F}$ and a \emph{%
given} warping family $\mathcal{H}$, the decomposition (\ref{eq:decomp}) 
\emph{is} identifiable (except for the usual indeterminacy on the sign of
the $\xi _{k} $s). But different combinations of templates and warping
models can give rise to essentially equivalent fits. The usual example is
the random shift: if $f_{i}(t)=\mu (t-\tau _{i})$, a simple Taylor
approximation yields $f_{i}(t)\approx \mu (t)-\tau _{i}\mu ^{\prime }(t)$,
so the $f_{i}$s could be modeled by a one-amplitude-component model without
warping just as well. Therefore, when we talk about \textquotedblleft
the\textquotedblright\ amplitude component and \textquotedblleft
the\textquotedblright\ warping component in this paper, it is always in the
context of a specific pair $(\mathcal{F},\mathcal{H})$.

Going back to the original problem: putting together (\ref{eq:raw-data}), (%
\ref{eq:decomp}), $\mathcal{F}$ and $\mathcal{H}$, and assuming the $%
\varepsilon _{ij}$s are i.i.d.~$N(0,\sigma ^{2})$, we obtain the following
random-effects model for the raw data $\mathbf{y}_{i}=(y_{i1},\ldots
,y_{im_{i}})$: 
\begin{eqnarray}
\mathbf{y}_{i}|(\mathbf{\theta }_{i},\mathbf{z}_{i}) &\sim &N_{m_{i}}\{%
\mathbf{\Phi }_{i}(\mathbf{\theta }_{i})(\mathbf{a}+\mathbf{Cz}_{i}),\sigma
^{2}\mathbf{I}_{m_{i}}\},  \label{eq:normal_model} \\
\mathbf{\theta }_{i} &\sim &N_{r}(\mathbf{\theta }_{0},\mathbf{\Sigma }), 
\nonumber \\
\mathbf{z}_{i} &\sim &N_{p}(\mathbf{0},\mathbf{\Lambda }),  \nonumber
\end{eqnarray}%
with $\mathbf{C}=[\mathbf{c}_{1},\ldots ,\mathbf{c}_{p}]$ and $\mathbf{\Phi }%
_{i}(\mathbf{\theta }_{i})$ the $m_{i}\times q$ matrix with elements $[%
\mathbf{\Phi }_{i}(\mathbf{\theta }_{i})]_{jl}=\phi _{l}\{g^{-1}(t_{ij},%
\mathbf{\theta }_{i})\}$ (the inverse of $g$ is taken with respect to the
variable $t$ for each $\mathbf{\theta }_{i}$.) The model parameters $\mathbf{%
a}$, $\mathbf{C}$, $\sigma ^{2}$, $\mathbf{\Sigma }$ and $\mathbf{\Lambda }$
are estimated by maximum likelihood using the EM algorithm. A drawback of
this approach is that it was developed for sparse and irregular time grids,
and it becomes infeasible for large $m_{i}$s; therefore we down-sampled the
curves so that $m_{i}=30$ for all $i$. Some high-definition features were
lost, but the main peaks are still clearly visible in Figure \ref%
{fig:data_curves}.

The random-effect approach to warping described in this section is still
unpublished for univariate samples, but a similar approach in the context of
functional regression is described in Gervini (2012), where the interested
reader may find more technical details.

\section{Results}

\FRAME{ftbpFU}{6.5795in}{2.0029in}{0pt}{\Qcb{Amplitude principal components.
Mean function (solid line), mean plus principal component (dash-dot line),
and mean minus principal component (dotted line), for first [(a)] and second
[(b)] principal component.}}{\Qlb{fig:pcs}}{pcs.eps}{\special{language
"Scientific Word";type "GRAPHIC";maintain-aspect-ratio TRUE;display
"USEDEF";valid_file "F";width 6.5795in;height 2.0029in;depth
0pt;original-width 10.92in;original-height 3.2906in;cropleft "0";croptop
"1";cropright "1";cropbottom "0";filename '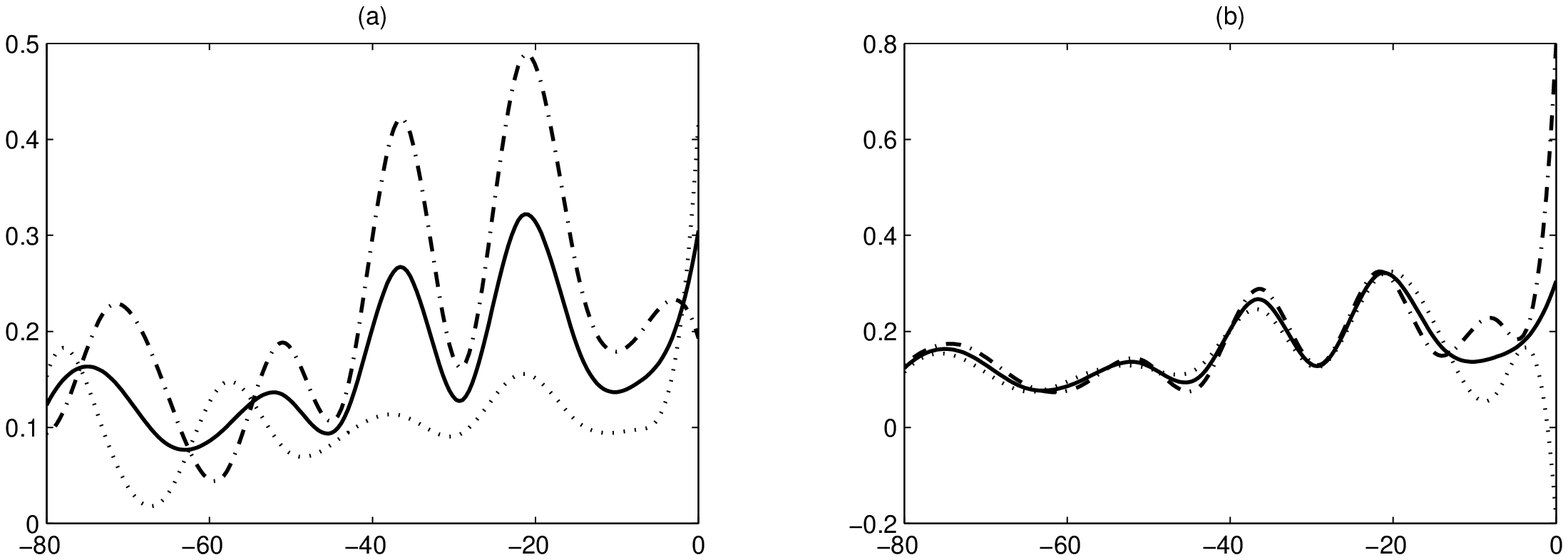';file-properties "XNPEU";}%
}

We fitted several models with warping knots $\mathbf{\tau }%
_{0}=(-60,-40,-20) $ and different numbers of amplitude components $p$
ranging from 0 (mean-only model) to 5. We used cubic B-splines with 10
equispaced knots for $\mu $ and the $\xi _{k}$s. The warped functions for $%
p=2$ are shown in Figures \ref{fig:data_curves}(b) and \ref{fig:data_curves}%
(d). Plots of $\hat{\mu}$ plus/minus $\hat{\xi}_{1}$ and $\hat{\xi}_{2}$ are
shown in Figure \ref{fig:pcs}. The first principal component is mostly
associated with amplitude variation at the syphon peaks, while the second
component is mostly associated with amplitude variation at the origin. Can
they be used to discriminate patients with aneurysms at or after the
terminal bifurcation of the internal carotid artery (the \textquotedblleft
upper\textquotedblright\ group) from patients with no-aneurysms or with
aneurysms along the carotid artery (the \textquotedblleft
lower\textquotedblright\ group)?

To answer this question we first tried logistic discrimination based on the
registered curves $\tilde{f}_{1},\ldots ,\tilde{f}_{n}$. Introducing a
binary variable $y$, with $y_{i}=1$ indicating the \textquotedblleft
upper\textquotedblright\ group and $y_{i}=0$ the rest of the patients, the
logistic model assumes that 
\begin{equation}
p(y_{i}=1|\tilde{f}_{i})=\func{logist}\left[ \alpha +\int_{I}\beta (t)\{%
\tilde{f}_{i}(t)-\mu (t)\}dt\right]  \label{eq:logistic}
\end{equation}%
for parameters $\alpha \in \mathbb{R}$ and $\beta \in \mathbb{L}^{2}(I)$.
Without loss of generality we can assume $\beta \in \mathrm{span}\{\xi
_{1},\ldots ,\xi _{p}\}$, since in view of (\ref{eq:KL}) the part of $\beta $
orthogonal to $\mathrm{span}\{\xi _{1},\ldots ,\xi _{p}\}$ will also be
orthogonal to $\tilde{f}_{i}-\mu $. Then we have $\beta
(t)=\sum_{k=1}^{p}b_{k}\xi _{k}(t)$ and we can re-write (\ref{eq:logistic})
as 
\begin{equation}
p(y_{i}=1|\tilde{f}_{i})=\func{logist}\left( \alpha +\mathbf{b}^{T}\mathbf{z}%
_{i}\right) ,  \label{eq:logistic_scores}
\end{equation}%
which is just a common multivariate logistic model. The parameters $\alpha $
and $\mathbf{b}$ were estimated by conditional maximum likelihood, as usual.
The crossvalidated misclassification rates for each $p$ are given in Table %
\ref{tab:CMR} (first column). The lowest one is attained at $p=4$, but in
the interest of parsimony we choose the second-best, the two-component
model, for which the misclassification rate is only slightly larger at
38.5\%.

This high misclassification rate is disappointing, and we wonder if the
warping process may not contain additional information that could be useful
for discrimination. An easy way to answer this question is to augment model (%
\ref{eq:logistic_scores}) with the $\mathbf{\tau }_{i}$s and assume that 
\begin{equation}
p(y_{i}=1|f_{i})=\func{logist}\left( \alpha +\mathbf{b}^{T}\mathbf{z}_{i}+%
\mathbf{d}^{T}\mathbf{\tau }_{i}\right) .  \label{eq:augmented_logistic}
\end{equation}%
Estimating the parameters by conditional maximum likelihood as before, the
crossvalidated misclassification rates we now obtain (Table \ref{tab:CMR},
second column) are considerably lower, in particular for the optimal
two-component model, which is 24.6\%. The parameter estimators are $\mathbf{%
\hat{b}}=(-8.12,-6.43)$ and $\mathbf{\hat{d}}=(-.15,.22,.27)$. The sign of $%
\mathbf{\hat{b}}$ indicates that the probability of being in the
\textquotedblleft upper\textquotedblright\ group decreases as the height of
the peaks at $t=-40$, $t=-20$ and $t=0$ increases (this is somewhat visible
to the naked eye in Figure \ref{fig:data_curves}(b) and \ref{fig:data_curves}%
(d).) The signs of the last two coefficients of $\mathbf{\hat{d}}$ also
indicate that for patients in the \textquotedblleft upper\textquotedblright\
group the peaks at $t=-40$ and $t=-20$ tend to occur closer to the origin; a
caveat is that this could be an artifact of the image-capturing process and
not a feature of artery shape, although the negative sign of $\hat{d}_{1}$
seems to rule this out (because, if the whole curve had been shifted, $\hat{d%
}_{1}$ would also be positive). Either way, this example shows that the
warping process sometimes does contain useful information for classification
and discrimination that should not be neglected.

\begin{table}[tbp] \centering%
\begin{tabular}{ccc}
\hline\hline
& \multicolumn{2}{c}{CMRs (\%)} \\ \cline{2-3}
$p$ & without $\mathbf{\tau }$s & with $\mathbf{\tau }$s \\ \hline
0 & --- & 41.5 \\ 
1 & 49.2 & 35.4 \\ 
2 & 38.5 & 24.6 \\ 
3 & 47.7 & 35.4 \\ 
4 & 36.9 & 35.4 \\ 
5 & 58.5 & 46.1 \\ \hline\hline
\end{tabular}%
\caption{Crossvalidated misclassification rates for models with $p$ amplitude components,
with and without warping parameters included in the model.}\label{tab:CMR}%
\end{table}%

There are a number of ways in which this analysis could be refined. For
example, instead of the two-step process followed above, where estimation of
amplitude principal components and warping functions is done separately from
discrimination, both steps could be brought together by maximizing the
likelihood of model (\ref{eq:augmented_logistic}) instead of (\ref%
{eq:normal_model}). The principal components and warping functions thus
obtained would have been optimized for discrimination and may yield lower
misclassification rates than the two-step process; the author is currently
investigating this approach. The other important issue is the handling of
incomplete curves. The approach in this analysis was to down-sample the
curves and apply a likelihood-based method originally developed for sparsely
sampled curves, but in doing so, the sharpest peaks of the curves are dulled
or lost entirely; that did not matter much for these data, but in other
situations the impact may be significant. The existing registration methods
that handle densely sampled curves usually involve functional inner products
and norms that require computation of integrals over the whole range $I$,
which cannot be done with incomplete curves (not in an elegant way at least,
i.e.~avoiding artificial truncations or extrapolations). Finding a way
around this problem would be an interesting contribution to the registration
literature.

\section*{Acknowledgements}

This research was partially supported by NSF grant DMS 10-06281. The author
also thanks the Mathematical Biosciences Institute (MBI) for funding his
participation in the workshop.

\section*{References}

\begin{description}
\item Fritsch, F.N.~and Carlson, R.E. (1980). Monotone piecewise cubic
interpolation. \emph{SIAM J. Numer. Anal. }\textbf{17} 238--246.

\item Gervini, D. (2012). Warped functional regression. \emph{ArXiv
1203.1975.}

\item Piccinelli, M., Bacigaluppi, S., Boccardi, E., Ene-Iordache, B.,
Remuzzi, A., Veneziani, A., and Antiga, L. (2011). Influence of internal
carotid artery geometry on aneurysm location and orientation: a
computational geometry study. \emph{Neurosurgery} \textbf{68 }1270--1285.

\item Sangalli, L.M., Secchi, P., and Vantini, S. (2013). AneuRisk65. \emph{%
Special Section, Electronic Journal of Statistics.}
\end{description}

\end{document}